\documentclass[12pt]{article}
\pdfoutput=1

\usepackage{epsfig}
\usepackage{psfrag}
\usepackage{latexsym}
\usepackage{amsmath}
\usepackage{amssymb}
\usepackage{amsfonts}
\usepackage{bbold}
\usepackage{graphicx}
\usepackage{cite}
\usepackage{pstricks}
\usepackage{inputenc}
\usepackage{verbatim}
\usepackage{slashed}
\usepackage{array}
\newpsobject{grilla}{psgrid}{subgriddiv=1,griddots=10,gridlabels=6pt}

\def\eq#1{{eq. (\ref{#1})}}

\def\vev#1{\left\langle #1\right\rangle}

\def\Im{\mbox{Im}\,}
\def\Re{\mbox{Re}\,}

\def\hbar{\hspace{0pt}\raisebox{1pt}{$-$} \hspace{-7pt} h}

\newcommand{\be}{\begin{equation}}
\newcommand{\ee}{\end{equation}}
\newcommand{\bea}{\begin{eqnarray}}
\newcommand{\eea}{\end{eqnarray}}
\newcommand{\nn}{\nonumber}

\def\5{\overline 5}

\renewcommand{\Re}{\mbox{Re}}
\renewcommand{\Im}{\mbox{Im}}

\let\vev\VEV

\def\roughly#1{\mathrel{\raise.3ex\hbox{$#1$\kern-.75em
      \lower1ex\hbox{$\sim$}}}} 

\def\e6{E(6)}
\def\321{$SU(3)_{c}\otimes SU(2)_L \otimes U(1)$}
\def\10{SO(10)}

\def\422{SU(4) $\otimes$ SU(2) $\otimes$ SU(2)}

\newcommand{\beq}{\begin{equation}}
\newcommand{\eeq}{\end{equation}}
\newcommand{\bac}{\beq\begin{array}}
\newcommand{\eac}{\end{array}\eeq}
\newcommand{\ba}{\begin{array}}
\newcommand{\ea}{\end{array}}
\newcommand{\esp}{\end{split}}
\newcommand{\bsp}{\begin{split}}

\newcommand{\bmat}{\begin{pmatrix}}
\newcommand{\emat}{\end{pmatrix}}
\def\author#1{\begin{center} #1\end{center}}

\begin{document}
\begin{titlepage}
\vspace*{-1cm}
\phantom{hep-ph/***}

\hfil{NIKHEF 2010-047}
\hfill{TUM-HEP-782/10}
\hfill{DFPD/2010/TH/20}

\vskip 1.3cm
\begin{center}
\mathversion{bold}
{\Large \bf  Constraining Flavour Symmetries At The EW Scale II:\\The Fermion Processes}
\mathversion{normal}
\end{center}

\vskip 0.5  cm

\begin{center}
 {\large Reinier de Adelhart Toorop}~$^{a)}$\footnote{e-mail address: reinier.de.adelhart.toorop@nikhef.nl},
{\large Federica Bazzocchi}~$^{b)}$\footnote{e-mail address: fbazzoc@few.vu.nl},
 \vskip 0.1cm
{\large Luca Merlo}~$^{c)}$\footnote{e-mail address: luca.merlo@ph.tum.de}
and {\large Alessio Paris}~$^{d)}$\footnote{e-mail address: alessio.paris@pd.infn.it}
\\
\vskip .1cm
$^{a)}$~Nikhef Theory Group, \\
Science Park 105, 1098 XG, Amsterdam, The Netherlands
\\
\vskip .1cm
$^{b)}$~Department of Physics and Astronomy, Vrije Universiteit Amsterdam,\\
1081 HV Amsterdam, The Netherlands
\\
\vskip .1cm
$^{c)}$Physik-Department, Technische Universit\"at M\"unchen\\
James-Franck-Str. 1, D-85748 Garching, Germany\\
TUM Institute of Advanced Study\\
Lichtenbergstr. 2a, D-85748 Garching, Germany
\vskip .1cm
$^{d)}$~Dipartimento di Fisica `G.~Galilei', Universit\`a di Padova\\
INFN, Sezione di Padova, Via Marzolo~8, I-35131 Padua, Italy\\
\end{center}

\vskip 0.3cm

\begin{abstract}
We study the set of models in which the Standard Model symmetry is extended with the flavour group $A_4$ and there are three copies of the Standard Model Higgs that transform as a triplet under this group. In this setup,  new channels for flavour violating processes can be studied once the $A_4$ representations of the fermions in the theory are given. We show that it is of great importance to take these constraints into account as they can put severe constraints on the viability of flavour models.
\end{abstract}

\end{titlepage}

\vskip2truecm
%
%
\newpage

\section{Introduction}

Discrete flavour symmetries  have been quite popular in the particle physics community since the discovery of neutrino oscillations  and the observation that the related neutrino mixing matrix may show interesting structures.

In a recent paper \cite{ABMP:Constraining1}, we discussed a specific class of these models. We observed that many models in the literature are well able to explain structures in the fermion mass sector, but that they are often rather involved, having a very large new physics sector. We noticed that it is often needed to break the flavour symmetry in different ways for different sectors. For instance the symmetry group $A_4$ is often invoked to reproduce the so-called Tri-Bimaximal mixing pattern. This can be successfully obtained if we assume that the lepton mass terms break $A_4$ down to two of its maximal subgroups: $Z_3$ in the charged lepton sector and $Z_2$ in the neutrino sector. However, breaking a flavour symmetry in more than one direction is highly non-trivial. In \cite{ABMP:Constraining1}, we mentioned that so far only a few theoretical techniques are known, all of which depend on supersymmetry or extra dimensions. For a review of the use of discrete flavour symmetries, see \cite{AF:ReviewDiscreteSyms}.

We discussed the possibility of having only one breaking direction in flavour space. In that case, one can build relatively simple models, in which it is not needed to introduce separate \textit{flavons}, scalar fields that are charged under the flavour symmetry and are responsible for its breaking. Instead, the ordinary Standard Model (SM) Higgs field can take the role of the flavons by being in a non-trivial representation of the flavour symmetry \cite{MR:A4EWscale,MP:A4EWscale,LK:A4EWscale,HMPV:DiscreteDM,MMP:DarkMatterA4}. A disadvantage is that the models are less natural: without a high energy theory, the parameters should be of the same order of magnitude, but here
large hierarchies are necessary. Furthermore special mixing patterns such as the Tri-Bimaximal mixing are only possible for fine-tuned and ad-hoc values of the parameters of the model.

We discussed in some detail a setup where the flavour symmetry group is $A_4$ and where there are three copies of the SM Higgs field in the triplet representation of the group. We found that the $A_4$ symmetric Higgs potential allows three different solutions with real vacuum expectation values (vevs), as well as two solutions in which some of the Higgs fields take complex vevs.

We analysed the spectrum of the Higgs bosons in these cases. In general, we expect that three complex Higgs fields should give rise to five neutral Higgs bosons (and one Goldstone boson) and two charged Higgs bosons (and again one Goldstone boson). We found that in some of these cases, there are extra, unwanted Goldstone bosons or very light scalars and that it might be impossible to have all masses real.
In this paper, we will show that it is possible to solve these problems by adding a small term to the potential that softly breaks the $A_4$ potential.

We will also discuss the interactions between the new Higgs bosons and the fermions in the theory. When more than one Higgs boson couples to all the fermions, the fermion-Higgs interaction matrix generally is no longer diagonal and flavour violating processes can be mediated by the Higgs scalars and pseudoscalars. In these models, there are more channels available for rare fermion decays and meson oscillations. Experimental data place stringent bounds on the masses of the Higgses. We note that these bounds are dependent on details of the model, such as the $A_4$ representations of the fermions and were therefore not included in \cite{ABMP:Constraining1}.

After a general analysis of the flavour violating processes, we will apply this to three models from the literature. Ma and Rajasekaran \cite{MR:A4EWscale} use the alignment ($v$, $v$, $v$) and focus on the lepton sector, even if the possible extension to quark is sketched. Morisi and Peinaldo \cite{MP:A4EWscale} and Lavoura and Kuhbock \cite{LK:A4EWscale} discuss models where the Higgses are alligned as $(v e^{i \omega}, v e^{-i \omega}, r v)$ (or a permutation thereof), with the first paper focusing on leptons and the second on quarks.

We will see that the model of Ma and Rajasekaran (Model 1) is quite robust under the constraints from the Higgs sector and flavour violating processes, while the models of Morisi and Peinaldo (Model 2) and Lavoura and Kuhbock (Model 3) are strongly affected by the new constraints.

%
%
\mathversion{bold}
\section{The $A_4$ Invariant Potential And Soft $A_4$ Breaking}
\mathversion{normal}
\label{sec:pot}

The most general scalar potential invariant under the symmetry $A_4$ and including only an $A_4$ triplet Higgs $\Phi_a$ is
\bea
\label{A4pot}
V[ \Phi_a]&=& \mu^2 (\Phi_1^\dag \Phi_1+ \Phi_2^\dag \Phi_2+ \Phi_3^\dag \Phi_3)+ \lambda_1 (\Phi_1^\dag \Phi_1+ \Phi_2^\dag \Phi_2+ \Phi_3^\dag \Phi_3)^2 \nn\\
&+& \lambda_3 (\Phi_1^\dag\Phi_1 \Phi_2^\dag\Phi_2+ \Phi_1^\dag\Phi_1\Phi_3^\dag \Phi_3+ \Phi_2^\dag\Phi_2\Phi_3^\dag \Phi_3) \\
&+& \lambda_4 (\Phi_1^\dag \Phi_2 \Phi_2^\dag \Phi_1 +\Phi_1^\dag \Phi_3 \Phi_3^\dag \Phi_1+ \Phi_2^\dag \Phi_3 \Phi_3^\dag \Phi_2)\nn\\
&+&\frac{\lambda_5}{2}\bigg[ e^{i \epsilon}  [ (\Phi_1^\dag \Phi_2)^2+(\Phi_2^\dag \Phi_3)^2+(\Phi_3^\dag \Phi_1)^2]+  e^{-i \epsilon}   [(\Phi_2^\dag \Phi_1)^2+(\Phi_3^\dag \Phi_2)^2+(\Phi_1^\dag \Phi_3)^2] \bigg]\nn \,,
\eea
In   \cite{ABMP:Constraining1} we searched for the minima of the potential and studied the spectra they produce.
We remind here that when the flavour and the Electroweak (EW) symmetry are broken through the Higgs vevs, the Higgs fields can be written as
\be
\label{Higgsdecomp}
\Phi_a =
\frac{1}{\sqrt{2}}
\bmat
\Re \; \Phi^{1}_a + i \Im \; \Phi^{1}_a \\
\Re \; \Phi^{0}_a + i \Im \; \Phi^{0}_a
\emat
\rightarrow \frac{1}{\sqrt{2}}
\bmat
\Re \; \phi_a^1 + i \, \Im \; \phi_a^1 \\
v_a e^{i \omega_a} + \Re \ \phi_a^0 + i \; \Im \, \phi_a^0
\emat.
\ee
The minima can be divided in two distinct classes, real and complex. To the first class belong the cases $(v,v,v)$ and $(v,0,0)$, that preserve the subgroup $Z_3$ and $Z_2$ of $A_4$ respectively. The alignment $(v_1,v_2,v_3)$ breaks $A_4$ completely and can only be obtained if there are fixed relations among the parameters $\lambda_i$ of the potential. This enlarges
the symmetry and gives rise to  unwanted extra-Goldstone bosons.

In the first class we have $(v e^{i \omega},v,0)$, which can only be safe if $\lambda_5 \neq 0$. The other possibility considered is the vacuum $(e^{i \omega},e^{-i \omega},r)v_w/\sqrt{2+r^2}$ , of which the limit
with $r$ very large is the most interesting choice and will be discussed in details in this work.
In fact, in one of the following section we will analyze the predictions in the flavour sector of three specific models, in which the chosen vacua are $(v,v,v)$, $(v e^{i \omega}, v e^{-i \omega}, r v )$
and a permutation of the latter. Regarding the last vev, in our previous paper we also stressed that some very light Higgs masses are expected.
To avoid this feature, which is potentially in contrast with the current limits on flavour violation, we add
soft breaking terms to eq. (\ref{A4pot}) in the form
\be
\label{softpot}
V_{A_4soft}= v_w^2 \frac{m}{2} (\phi_1^\dag \phi_2+\phi_2^\dag \phi_1)+ v_w^2 \frac{n}{2} (\phi_2^\dag \phi_3+\phi_3^\dag \phi_2) + v_w^2 \frac{k}{2} (\phi_1^\dag \phi_3+\phi_3^\dag \phi_1)\,,
\ee
where $m,n,k$ are adimensional parameters that should presumably be smaller than one. Notice that the chosen $V_{A_4soft}$ is not the most general one but it prevents accidental extra $U(1)$ factor to appear.

%
%
\section{General Analysis Of The Higgs-Fermion Interactions}
\label{sec:interac}

We consider three Higgs fields $\Phi_a$ with hypercharge $+1/2$ transforming as doublets under $SU(2)_L$ and as a triplet under a generic flavour symmetry  $ G_f$, eventually to be identified with $A_4$.
Each $\Phi_a$ will couple to the three  fermion families  according to the group rules. Without specifying the flavour group $G_f$ and the fermion representations under it, in general $\Phi_a$ will couple to fermions through a given $Y^a_{ij}$
\begin{equation}
 \mathcal{L}_Y  =  \left(Y^d_{ija} \overline{Q}_{L i} d_{R j} \Phi_a  + Y^u_{ija} \overline{Q}_{L i} u_{R j} \Phi_a^\dagger\right)+\left(d \leftrightarrow e\right)  + h.c.
\end{equation}
where $i$ and $j$ are fermion family indices and $a$ is the Higgs triplet index. Notice that, in order to keep the formulae compact, we simply use $d \leftrightarrow e$ to indicate that similar Yukawa terms are present in which down quarks are substituted by charged leptons. Without specifying any high-energy explanation for the neutrino masses, we consider the low-energy effective Weinberg operator: this term generates the neutrino masses and it has been already discussed in the models we will analyze in the next sections.
After EW symmetry breaking according to eq. (\ref{Higgsdecomp}), the part of the Lagrangian including neutral Higgs fields becomes
\begin{equation}
\bsp
\mathcal{L}_{Y,n} = &\left(Y^d_{ija} \overline{Q}^d_{L i} d_{R j} \frac{v_a e^{i \omega_a}}{\sqrt{2}} + Y^d_{ija} \overline{Q}^d_{L i} d_{R j} \frac{1}{\sqrt{2}} (\Re \ \phi_a^0 + i \; \Im \, \phi_a^0)+\right.\\[2mm]
&+\left.Y^u_{ija} \overline{Q}^u_{L i} u_{R j} \frac{v_a e^{-i \omega_a}}{\sqrt{2}}+ Y^u_{ija} \overline{Q}^u_{L i} u_{R j} \frac{1}{\sqrt{2}} (\Re \ \phi_a^0 - i \; \Im \, \phi_a^0) \right)+\\[2mm]
&+\left( d \leftrightarrow e\right)  + h.c.,
\end{split}
\end{equation}
while the part with the charged Higgs is
\begin{equation}
\mathcal{L}_{Y,ch} =\left(Y^d_{ija} \overline{Q}^u_{L i} d_{R j} \Phi^1_a-Y^u_{ija} \overline{Q}^d_{L i} u_{R j} (\Phi^1_a)^*\right)+ \left(d \leftrightarrow e\right) + h.c.
\end{equation}
Now we move to the mass basis of fermions through the transformations:
\beq
\overline{Q}_{Li}^d = \overline{\hat{Q}}_{Lr}^d V_{Lri}^{d\dagger}\,,\qquad  \qquad  d_{R j} = V_{Rjs}^d \hat{d}_{R s}\,, \eeq
and in analogous way for all the other particles.
The neutral  and the charged Higgs fields are also rotated into the mass basis:
\beq
\label{defS}
\bmat h_1 \\ \vdots \\ h_5 \\ \pi^0 \emat = U \bmat \Re \, \phi_1^0 \\ \vdots \\ \Re \, \phi_3^0 \\ \Im \, \phi_1^0 \\ \vdots \\ \Im \, \phi_3^0 \emat\,,\qquad\qquad \bmat \hat{H}^+_{1} \\ \hat{H}^+_{2} \\ \pi^+  \emat = S\, \bmat \Phi^1_1 \\ \Phi^1_2 \\ \Phi^1_3 \emat\,,
\eeq
where
$\pi^\pm$ and $\pi^0$ are the Goldstone bosons associated to EW symmetry breaking.

In the mass basis the part of the Lagrangian which includes the neutral Higgs becomes
\beq
\begin{split}
 \mathcal{L}_{Y,n} =& \left(\overline{\hat{d}}_r M_{(r)}^d \frac{1 + \gamma_5}{2} \hat{d}_r
+ \overline{\hat{d}}_{r} (R^d)_{r s}^\alpha  h_\alpha \frac{1 + \gamma_5}{2} \hat{d}_{s}+ \overline{\hat{u}}_r M_{(r)}^u \frac{1 + \gamma_5}{2} \hat{u}_r
+ \overline{\hat{u}}_{r} (R^u)_{r s}^\alpha  h_\alpha \frac{1 + \gamma_5}{2} \hat{u}_{s}\right) +\\[2mm]
&+\left(d \leftrightarrow e\right) + h.c.
\end{split}
\label{eq10}
\eeq
with
\beq
\ba{rcl}
 M_{ij}^{d,u} &=& V_{Lri}^{d,u \dagger}\left( \sum_a \dfrac{v_a}{\sqrt{2}}  Y_{i j a}^{d,u} \right) V_{Rjs}^{d,u}\,,\\[4mm]
 (R^d)_{r s}^\alpha&=&\left[ V_{Lri}^{d \dagger} \dfrac{1}{\sqrt{2}} (i U^{\dagger (a+3) \alpha} + U^\dagger{}^{a \alpha})  Y_{i j a}^{d}   V_{Rjs}^{d}\right]\,, \\[4mm]
(R^u)_{r s}^\alpha&=& \left[ V_{Lri}^{u \dagger} \dfrac{1}{\sqrt{2}} (-i U^{\dagger (a+3) \alpha} + U^\dagger{}^{a \alpha})  Y_{i j a}^{u}   V_{Rjs}^{u}\right]\,,
\ea
\eeq
and similarly for the leptons.
The interaction with the charged Higgs becomes
\beq
\mathcal{L}_{Y,ch} =\left(\overline{\hat{u}}_r (T^d)_{r s}^\beta \hat{H}^+_\beta \frac{1 + \gamma_5}{2}  \hat{d}_{s}
                    -\overline{\hat{d}}_r (T^u)_{r s}^\beta \hat{H}^-_\beta \frac{1 + \gamma_5}{2}  \hat{u}_{s}\right)+\left(d \leftrightarrow e\right) +h.c.
\eeq
where
\beq
(T^{d,u})_{r s}^\beta=\Big[ V_{Lri}^{d,u \dagger}S^\dagger{}^{b \beta}   Y^{d,u}_{ij b} V_{Rjs}^{d,u}\Big]\,
\eeq
and similarly for the leptons.
Expanding the hermitian conjugate, the Lagrangian can be written in a more compact form
\beq
\bsp
\mathcal{L}_Y  =& \left(\overline{\hat{d}}_r M_{(r)}^d  \hat{d}_r + \overline{\hat{d}}_{r} \big((I^d)_{r,s}^\alpha +\gamma_5 (J^d)_{r,s}^\alpha \big)  h_\alpha  \hat{d}_{s}\right.\\[2mm]
& + \overline{\hat{u}}_r M_{(r)}^u \hat{u}_r + \overline{\hat{u}}_{r} \big((I^u)_{r,s}^\alpha +\gamma_5 (J^u)_{r,s}^\alpha \big)  h_\alpha \hat{u}_{s} \\[2mm]
& \left.+\overline{\hat{u}}_r \big(F_{r,s}^\beta +\gamma_5 G_{r,s}^\beta \big) \hat{H}^+_\beta \hat{d}_{s} +\overline{\hat{d}}_r \big(F_{r,s}^{\beta *} -\gamma_5 G_{r,s}^{\beta *} \big) \hat{H}^-_\beta \hat{u}_{s}\right)+ \left(d \leftrightarrow e \right)\,,
\end{split}
\eeq
with the new coefficients defined in the following way:
\beq
\ba{ccl}
(I^{d,u})_{r,s}^\alpha &=& \dfrac{1}{2} \Big((R^{d,u})_{r s}^\alpha + ((R^{d,u})_{s r}^\alpha)^* \Big)\,, \\[2mm]
(J^{d,u})_{r,s}^\alpha &=& \dfrac{1}{2} \Big((R^{d,u})_{r s}^\alpha - ((R^{d,u})_{s r}^\alpha)^* \Big)\,, \\[2mm]
F_{r,s}^\beta &=& \dfrac{1}{2} \Big((T^{d})_{r s}^\beta)^* - ((T^{u})_{s r}^\beta)^* \Big)\,, \\[2mm]
G_{r,s}^\beta &=& \dfrac{1}{2} \Big((T^{d})_{r s}^\beta)^* +((T^{u})_{s r}^\beta)^* \Big)\,,
\ea
\eeq
and similarly for the leptons. As shown in the next section, the operators $I$, $J$, $F$ and $G$ determine whether flavour changing interactions are possible and what their strength is. Note that for particularly symmetric vevs of the Higgs fields, many of these operators are automatically zero, thus forbidding many flavour changing interactions or allowing them only if certain selection rules are met.

\subsection{Flavor Changing Interactions}
\label{sec:decay}

The interaction of fermions with the Higgs particles induces flavour violating processes in the lepton and quark sectors. In the first one, rare decays of muon and tau particles
into three leptons are allowed at tree-level, while processes as $l_i \rightarrow l_j \gamma$ take place through one-loop graphs. For the quarks the possibility of $\Delta F = 2$ meson-antimeson oscillations is considered.

\subsubsection*{The Processes $\mu^- \rightarrow e^- e^- e^+$ and $\tau^- \rightarrow \mu^- \mu^- e^+$ }

\begin{figure}[ht!]
\begin{center}
\includegraphics[height=5.5cm]{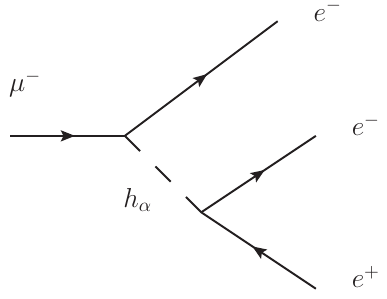} \qquad
\includegraphics[height =5.5cm]{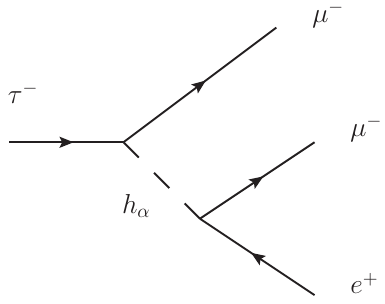}
\caption{\label{fig1}\it The decays $\mu^- \rightarrow e^+ e^- e^-$ (left) and  $\tau^- \rightarrow e^+ \mu^- \mu^-$ (right) can occur at tree level in our models.}
\end{center}
\end{figure}

We consider the decay of a muon into a positron and two electrons (fig. \ref{fig1} on the left). In the approximation of massless final states, the decay amplitude is written as
\beq
\Gamma(\mu\rightarrow e e \overline{e}) = \frac{m_\mu^5}{(4 \pi)^3\times 24} I_{\mu e e e},
\eeq
where the coefficient $I_{\mu e e e}$ is a combination of $I_{ij}$ and $J_{ij}$, that were defined in the previous section:
\be
I_{\mu e e e}
= \bigg|\sum_\alpha \frac{I_{\mu e}^\alpha I_{e e}^\alpha}{m_H^\alpha{}^2}\bigg|^2 \ + \ \bigg|\sum_\alpha \frac{J_{\mu e}^\alpha J_{e e}^\alpha}{m_H^\alpha{}^2}\bigg|^2  \ + \ \bigg|\sum_\alpha \frac{I_{\mu e}^\alpha J_{e e}^\alpha}{m_H^\alpha{}^2}\bigg|^2  \ + \ \bigg|\sum_\alpha \frac{J_{\mu e}^\alpha I_{e e}^\alpha}{m_H^\alpha{}^2}\bigg|^2.
\ee

The prediction for the corresponding branching ratio is then
\beq
\textrm{Br} (\mu\rightarrow e e \overline{e}) \approx \dfrac{\Gamma(\mu\rightarrow e e \overline{e})}{\Gamma(\mu\rightarrow e \overline{\nu}_e \nu_\mu )}= \dfrac{I_{\mu e e e}}{8 G_F^2},
\eeq
to be compared with the experimental value \cite{PDG2010} of $\textrm{Br} (\mu\rightarrow e e \overline{e})_{exp} = 1.0 \times 10^{-12}$.\\

The decay of a $\tau$ into two muons and a positron (fig. \ref{fig1} on the right) is generally less constrained that the decay of the muon in two electrons and a positron, but it is of interest in models where the latter process is prohibited by the symmetries. The calculation proceeds in an analogous way. In fact, the decay amplitude is now
\beq
\Gamma(\tau \rightarrow \overline{e} \mu \mu) = \frac{m_\tau^5}{(4 \pi)^3\times 24} I_{\tau \mu \mu e},
\eeq
where the coefficient is now given by the following expression:
\be
I_{\tau \mu \mu e} =\bigg|\sum_\alpha \frac{I_{\tau \mu}^\alpha I_{e \mu}^\alpha}{m_H^\alpha{}^2}\bigg|^2 \ + \ \bigg|\sum_\alpha \frac{J_{\tau \mu}^\alpha J_{e \mu}^\alpha}{m_H^\alpha{}^2}\bigg|^2  \ + \ \bigg|\sum_\alpha \frac{I_{\tau \mu}^\alpha J_{e \mu}^\alpha}{m_H^\alpha{}^2}\bigg|^2  \ + \ \bigg|\sum_\alpha \frac{J_{\tau \mu}^\alpha I_{e \mu}^\alpha}{m_H^\alpha{}^2}\bigg|^2.
\ee
while the branching ratio becomes
\beq
\textrm{Br} (\tau \rightarrow \overline{e} \mu \mu) =0.17 \times  \dfrac{\Gamma(\tau \rightarrow \overline{e} \mu \mu)}{\Gamma(\tau\rightarrow \mu \overline{\nu}_\mu \nu_\tau )}=0.17 \times \dfrac{I_{\tau \mu \mu e}}{8 G_F^2},
\eeq
to be compared with the experimental limit $\textrm{Br} (\tau \rightarrow \overline{e} \mu \mu)_{exp}=2.3 \times 10^{-8}$ \cite{PDG2010}.

\subsubsection*{The process $\mu^- \rightarrow e^-  \gamma$}

\begin{figure}[ht!]
\begin{center}
\includegraphics[height=5.5cm]{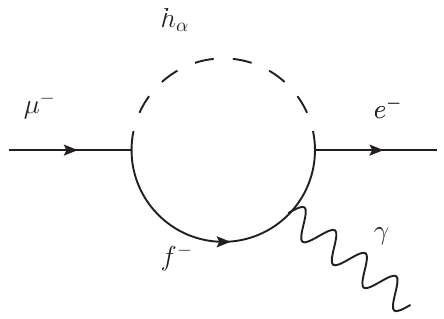}
\caption{\label{fig2}\it The decays $\mu^- \rightarrow e^- \gamma$ proceeds at one loop in our models, but can be much larger than in the Standard Model, where a GIM-like cancellation occurs.}
\end{center}
\end{figure}

The relevant diagram for this process has one loop with a charged fermion and a neutral Higgs (see fig. \ref{fig2}). We consider the limit in which the Higgs is much heavier than the virtual fermion and the final electron is massless.
Under this assumption the decay amplitude becomes \cite{Lavoura:GenFormulae}
\be
\Gamma (\mu\rightarrow e  \gamma) = \frac{e^2 m_\mu^5}{6\times (16)^3 \pi^5} \left| \sum_{\alpha, f} \frac{(R^\alpha_{f e})^* R^\alpha_{f \mu}}{m_H^\alpha{}^2} \right|^2
\ee
and the branching ratio is
\be
\textrm{Br} (\mu\rightarrow e  \gamma) =\frac{\Gamma (\mu \rightarrow e \gamma)}{\Gamma (\mu \rightarrow e  \nu  \overline{\nu})}=
\frac{\alpha_{em}}{32 \pi G_F^2}\left| \sum_{\alpha, f} \frac{(R^\alpha_{f e})^* R^\alpha_{f \mu}}{m_H^\alpha{}^2} \right|^2
\ee
to be compared with the current \cite{MEGA} (future \cite{MEG}) experimental bound $\textrm{Br}(\mu\rightarrow e \gamma)_{exp} =10^{-11}\; (10^{-13})$.

\subsubsection*{Meson oscillations}
\label{sec:osc}

\begin{figure}[ht!]
\begin{center}
\includegraphics[height=5cm]{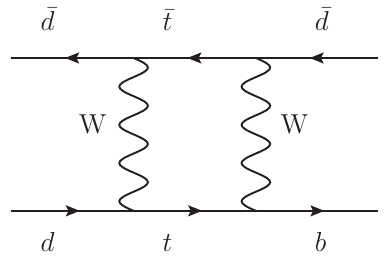} \qquad
\includegraphics[height=5cm]{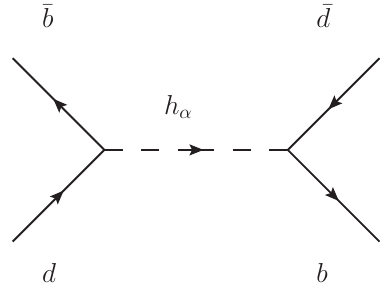}
\caption{\label{fig3} \it $B_d$-$\overline{B}_d$ oscillations take place via box diagrams in the Standard Model, but can proceed via tree-level Higgs exchange in our model.}
\end{center}
\end{figure}

Meson-antimeson oscillations are constrained to be generated by box processes in the SM (fig. \ref{fig3} on the left left), but in the presence of flavour violating Higgs couplings, they can also proceed via tree-level Higgs exchange.

For the mass splitting connected to $F^0 - \overline{F}^0$ oscillations \cite{ARS:Phem2HDM, Wells:HiggsLectures}, we find
\be
\Delta M_F = B_F^2 \ f_F^2 \ M_F \sum_\alpha \bigg[ \frac{1}{m_H^\alpha{}^2} \Big( |I^\alpha_{rs}|^2 \Big(\frac{1}{6} + \frac{1}{6} \frac{M_F^2}{(m_r + m_s)^2} \Big) +  |J^\alpha_{rs}|^2 \Big(\frac{1}{6} + \frac{11}{6} \frac{M_F^2}{(m_r + m_s)^2} \Big) \Big) \bigg].
\ee
Here, $M_F$ is the mass of the meson, $f_F$ is its decay constant and $B_F$ are recalibration constants of order 1, related to vacuum insertion formalism. Lastly, $m_r$ and $m_s$ are the masses of the quarks of which the meson is build, i.e. $rs = bd, \, bs, \, ds$ stands for $B_d, \, B_s$ and $K^0$ respectively.
Recent experimental values for the meson parameters, including $\Delta M_F$ that should be reproduced by the model, are given in table \ref{table1}

\begin{table}[!h]
\begin{center}
\begin{tabular}{|c||c|c|c|c|c|}
\hline
Meson & $M_F$ (GeV) & $f_F$ (GeV) & $B_F$ & $\Delta M_F$  (GeV)\\
\hline
$B_d$ ($\overline{b} d$) & $5.2795$ & $0.1928 \pm 0.0099$ & $1.26 \pm 0.11 $ & $(3.337 \pm 0.006 ) \times 10^{-13}$ \\
$B_s$ ($\overline{b} s$) & $5.3664$ & $0.2388 \pm 0.0095$ & $1.33 \pm 0.06 $ & $(1.170 \pm 0.008 ) \times 10^{-11}$ \\
$K$ ($\overline{s} d$) & $0.497614$ & $0.1558 \pm 0.0017$ & $0.725 \pm 0.026 $ & $(3.500 \pm 0.006 ) \times 10^{-11}$ \\
\hline
\end{tabular}
\end{center}
\caption{\it Properties of neutral mesons \cite{BCGI:HiggsFCNC}.}
\label{table1}
\end{table}

%
%
\section{$A_4$ models for quark and/or lepton masses}
\label{sec:models}

In this section we will apply the general results about flavour violation to three specific models. After describing the main features of each model, plots of relevant
flavour violating processes are reported. {The points belonging to the plots are not chosen casually, but instead represent parts of the parameter
space that fulfill the tests in the Higgs sector, as performed in \cite{ABMP:Constraining1} (positiveness of mass eigenstates, perturbative unitary constraints, bounds from $Z$ and $W$ decays and oblique corrections).}

\subsection{Model 1}
\label{sec:Ma}

The aim of the Model 1 \cite{MR:A4EWscale} is to reproduce the lepton mixing parameters in the Tri-Bimaximal frame, although it is not possible without introducing hierarchies among the parameters. Quarks are briefly mentioned in the paper, but the bulk of the analysis is about the lepton sector.
The triplet $\Phi_a$ couples only to charged leptons  and the chosen vacuum alignment falls in the class $(v,v,v)$, with $v$ real.
The Yukawa matrices in this sector are
\be
Y_{i j 1}=
\left(
\begin{array}{ccc}
y_1& y_2& y_3\\
0& 0& 0\\
0& 0 & 0
\end{array}
\right),\quad
Y_{i j 2}=
\left(
\begin{array}{ccc}
0& 0& 0\\
y_1& \omega y_2& \omega^2 y_3\\
0& 0 & 0
\end{array}
\right),\quad
Y_{i j 3}=
\left(
\begin{array}{ccc}
0& 0& 0\\
0& 0 & 0\\
y_1& \omega^2 y_2& \omega y_3
\end{array}
\right)
\label{yukawa_ma}
\ee
After the diagonalization of the charged lepton mass matrix, it is straightforward to relate the coefficient $y_i$ to the mass eigenvalues:
\be
y_1=\frac{m_e}{\sqrt{3} v}, \quad y_2=\frac{m_\mu}{\sqrt{3}v}, \quad y_3=\frac{m_\tau}{\sqrt{3}v}.
\ee
Since the vevs of the scalar potential are real, and consequently CP conserving, the $U$ matrix that rotates the Higgs fields into the mass basis (see eq. \ref{defS}) is block diagonal. Neutrino masses are given through a low scale ($\sim$ TeV) type I See-Saw implemented by $3$ right handed neutrinos that transform as an $A_4$ triplet  and by  an $SU(2)_L$ doublet Higgs, $\eta$, singlet of $A_4$
\be
\label{etadecomp}
\eta =\bmat
 \eta^{1} \\
\eta^{0}
\emat =
\frac{1}{\sqrt{2}}
\bmat
\Re \; \eta^{1}+ i \Im \; \eta^{1} \\
\Re \; \eta^{0} + i \Im \; \eta^{0}
\emat \,.
\ee
Clearly $\eta$ participates to the scalar potential, thus the Model 1 presents a scalar sector less minimal of that studied in \cite{ABMP:Constraining1}. In this specific case the new scalar potential added to \eq{A4pot} is given by
\beq
\label{etasc}
\ba{ccl}
V_\eta &=& \mu^2_\eta (\eta^\dag \eta) +\lambda_\eta (\eta^\dag \eta)^2+ \lambda_{\eta\Phi} (\eta^\dag \eta)(\phi_1^\dag \phi_1+\phi_2^\dag \phi_2+\phi_3^\dag \phi_3)   \,,\nn\\
V_{\eta\,soft} &=& \mu^2_{\eta\Phi} \left[\eta^\dag(\phi_1+\phi_2+\phi_3)+(\phi_1^\dag+\phi_2^\dag+\phi_3^\dag)\eta\right]\,,
\ea
\eeq
where the $A_4$ soft breaking part $V_{\eta\,soft}$ is needed in order to avoid additional GBs. $V_{\eta\,soft} $ breaks $A_4$ but preserves its $Z_3$ subgroup\footnote{Notice that in the original model \cite{Ma:SeeSawNONP}, $\eta$ is carrying lepton number, which is explicitly broken by soft terms. This prevents the appearance of further GBs.}, thus the full potential may naturally realize the vacuum configuration
\beq
\vev{\Phi}\sim (v,v,v)\,, \qquad\qquad \vev{\eta^0}\sim u\,.
\eeq
Notice that $u$ is responsible for neutrino masses and in the original model \cite{Ma:SeeSawNONP} it has been assumed to be tiny, $u\ll v\sim v_w/\sqrt{3}$. This may be easily realized if $\mu^2_{\eta \Phi}\sim\mathcal{O}(u\,v_w)$.

We have already demonstrated in \cite{ABMP:Constraining1} that it is not necessary to set $\epsilon$ to zero in eq. (\ref{A4pot}) to get this particular vev, as is assumed in \cite{MR:A4EWscale}. Moreover, since $Z_3$ is preserved, the mass eigenstates of the triplet $\Phi_a$, 5 neutral and 2 charged, can be arranged in $Z_3$ representations, as discussed for the case $(v,v,v)$ in \cite{ABMP:Constraining1}: moving to this $Z_3$ basis, we denote the states as  $\varphi$, $\varphi^{\prime}$ and  $\varphi^{\prime\prime}$, transforming as $1$, $1'\sim\omega$ and $1''\sim\omega^2$ of $Z_3$, respectively. This setup has been discussed in the context of the lepton triality in \cite{Ma:A4EWscale}. Notice that only the state $\varphi$ develops a non-vanishing vev in the neutral direction, while the other two are inert scalars. Moreover, $\varphi$ behaves as the SM-Higgs and acquires the mass $m_{h_1}$ defined in \cite{ABMP:Constraining1}. Furthermore, the transformation properties of the additional scalar $\eta$ allow a mixing between $\varphi^0$ ($\varphi^1$) and $\eta^0$ ($\eta^1$), both behaving as the SM-Higgs. However, this mixing interaction, $iZ \eta^0\varphi^0+h.c.$, that was not present in \cite{ABMP:Constraining1}, is irrelevant for the scalar spectrum discussion, because the coupling is extremely small being suppressed by $\sim u$. As a result, the conclusions driven in \cite{ABMP:Constraining1} for the case $(v,v,v)$ apply also in this context.

The coupling of the Higgses $\varphi^{\prime0}$, $\varphi^{\prime\prime0}$ to fermions is purely flavour violating. This setup has striking effects on the lepton processes. In fact it was shown in \cite{FP:RareDecaysA4} that, when the $A_4$ symmetry is unbroken, only a limited number of processes is allowed and these either conserve flavour or satisfy the constraint $\Delta L_e \times \Delta L_\mu \times \Delta L_\tau = \pm 2$. The only source of symmetry breaking is the vev of the SM-like Higgs $\varphi^0$, which is flavour-conserving and thus not involved in the processes we are looking at. We conclude that all flavour violating processes should satisfy the selection rule. In particular this implies that the decays $\mu^- \rightarrow e^- e^- e^+$ and $\mu \rightarrow e\gamma$ are not allowed, in the latter case in contrast with what was reported in \cite{MR:A4EWscale}, but in agreement with the more recent \cite{Ma:DarkTB}.

\begin{figure}[th!]
\begin{center}
\includegraphics[width=11cm]{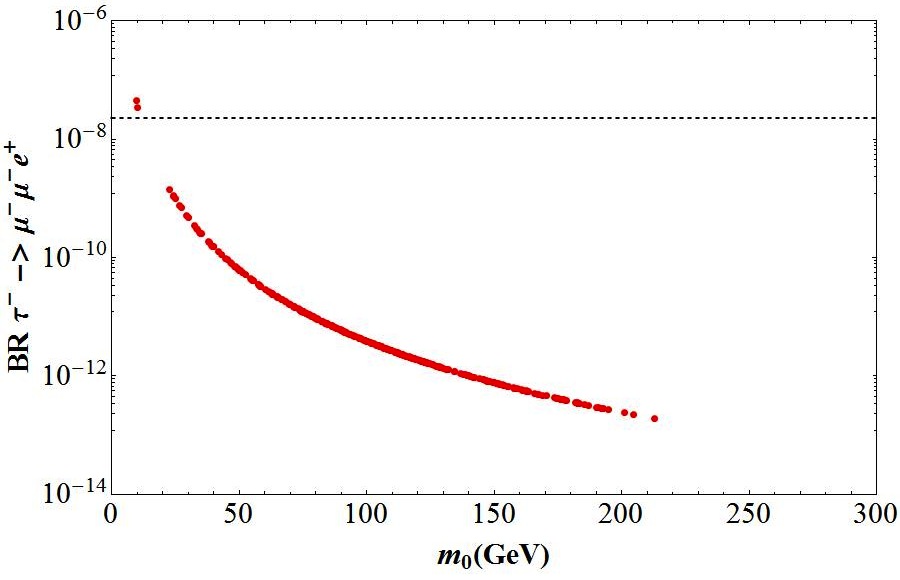}
\includegraphics[width=11cm]{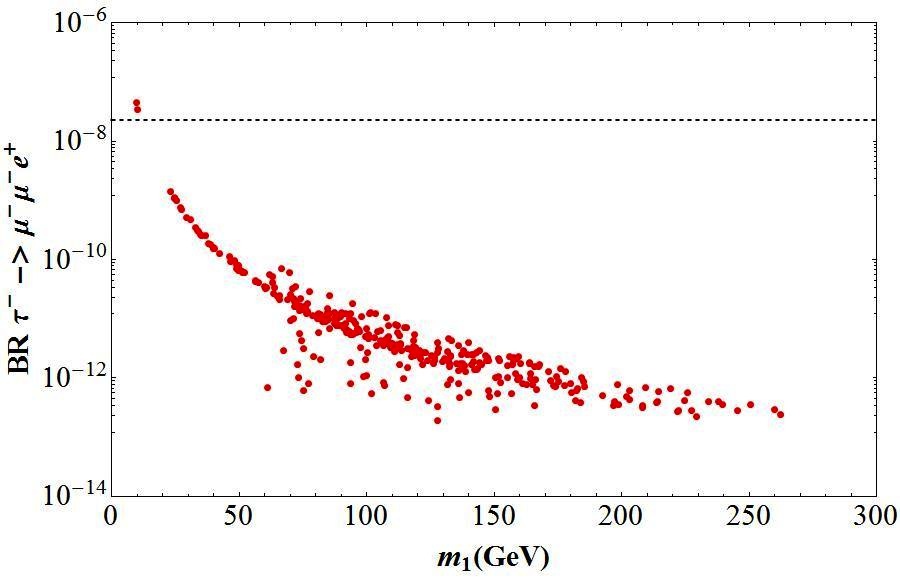}
\caption{\label{fig4} \it On the upper (lower) side, the branching ratio for the decay $\tau^- \rightarrow \mu^- \mu^- e^+$ as a function of the effective mass $m_0$ (the smallest mass $m_1$) in the situation where the parameter $\epsilon$ is zero. The horizontal line corresponds to the experimental upper bound.}
\end{center}
\end{figure}

Of the allowed processes, the less suppressed is $\tau^- \rightarrow \mu^- \mu^- e^+$, since its branching ratio is proportional to $m_\tau^2 m_\mu^2$. However, even this decay is very rare and below the experimental limit for most values of the Higgs masses. In the upper part of fig. \ref{fig4}, we plot the branching ratio for the decay against an effective mass defined as $m_0^{-2} = m_{h_A}^{-2} + m_{h_B}^{-2}$, where $A$ and $B$ are the two pairs of degenerate bosons. In the lower part, the same branching ratio against the mass of the lightest state, $m_1$. In both the plots, the parameter $\epsilon$ is set to zero, corresponding to the real Higgs potential discussed in \cite{MR:A4EWscale}. For the first picture, we reproduce the result of  \cite{MR:A4EWscale} that the branching ratio is proportional to $m_0^{-4}$. In the second one, this dependence is lost, even if we can see a similar behaviour. Once we take $\epsilon$ over the full range  [$0, 2 \pi$], we verified that the points cover a larger parameter space, but still concentrating around the previous points with $\epsilon=0$.

\begin{figure}[th!]
\begin{center}
\includegraphics[width=11cm]{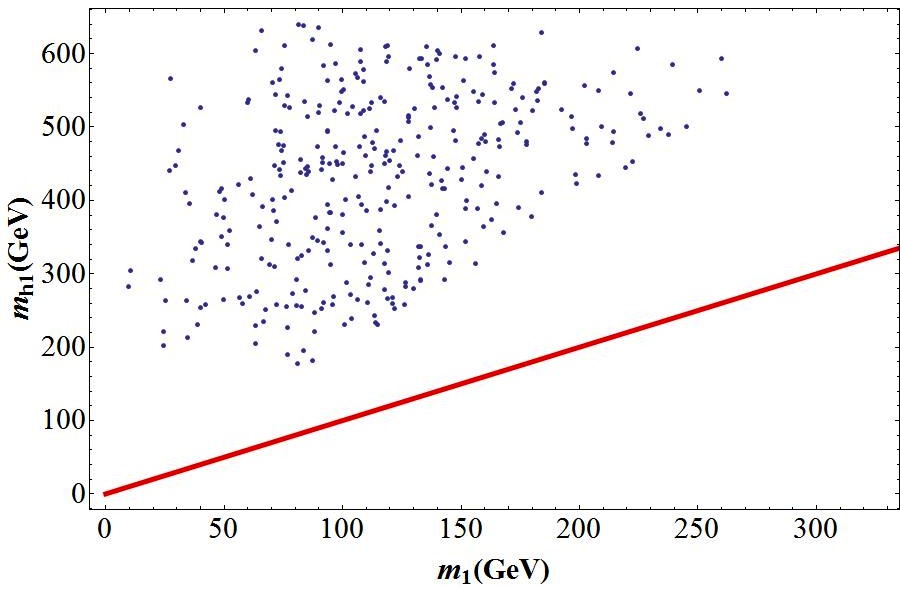}
\caption{\label{fig4b} \it The mass of the SM-Higgs $m_{h_1}$ against the smallest Higgs mass.}
\end{center}
\end{figure}

In fig. \ref{fig4b}, we show the masses of the SM-Higgs $\varphi^0$, $m_{h_1}$, against the mass of the lightest state $m_1$. A plot with the mass of the SM-Higgs $\eta^0$ against $m_1$ looks very similar to fig. \ref{fig4b}. All the points are above the diagonal and this corresponds to the fact that the SM-Higgses are always heavier than the lightest state. As already stated in \cite{ABMP:Constraining1}, in this situation, the standard upper bound of $194$ GeV at $99\%$ CL \cite{PDG2010} cannot apply due to the combined effect of the CP and $Z_3$ symmetries and the smallness of the $iZ \eta^0\varphi^0+h.c.$ coupling.

Finally, we can comment on the magnetic dipole moments, which could give interesting hints in this model. The discrepancy between the experimental measurement and the SM theoretical prediction of the magnetic dipole moment of the muon is usually a good test of flavour models \cite{FHLM:LFVinA4,FHM:Vacuum,FHLM:LFVinSUSYA4}, which could in principle provide new contributions. However, in this particular model it has already been discussed in \cite{MR:A4EWscale} that the non-SM contributions are negligible.

\subsection{Model 2}
\label{sec:Mor}

As in the previous section, the Model 2 \cite{MP:A4EWscale} deals only with the lepton sector, but the vacuum configuration used is different: here $(r,e^{i \omega},e^{-i \omega})v_w/\sqrt{2+r^2}$ is assumed, where $r$ is an adimensional quantity. The Yukawa texture in the charged lepton sector depends on two parameters:
\be
Y_{i j 1}=
\left(
\begin{array}{ccc}
0& 0& 0\\
0& 0 & y_1\\
0& y_2 & 0
\end{array}
\right),\quad
Y_{i j 2}=
\left(
\begin{array}{ccc}
0& 0& y_2\\
0& 0& 0\\
y_1& 0 & 0
\end{array}
\right),\quad
Y_{i j 3}=
\left(
\begin{array}{ccc}
0& y_1& 0\\
y_2& 0 & 0\\
0 & 0& 0
\end{array}
\right).
\label{yukawa_morisi}
\ee
In order to reproduce the masses of the leptons, $r\simeq 240$ is fit to the data and as a result the minimum of the scalar potential falls in the large $r$
scenario, as discussed in \cite{ABMP:Constraining1}. The final number of the parameters in this model is four, two coming from the Yukawas and two from the vacuum configuration.

\begin{figure}[h!]
\begin{center}
\includegraphics[width=10cm]{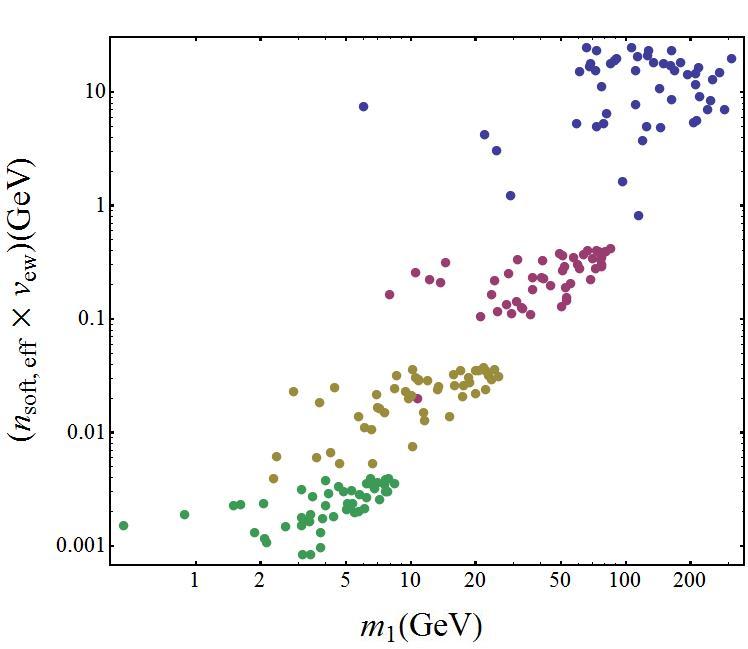}
\caption{\label{fig:soft}\it Correlation among the lightest Higgs mass and the soft breaking parameters. The different colours correspond to the
ranges that the individual parameters $m$, $n$ and $k$ are in, respectively $(0-10^{-4})$, $(0-10^{-3})$, $(0-10^{-2})$ and $(0-10^{-1})$}
\end{center}
\end{figure}

We have studied this vev alignment in \cite{ABMP:Constraining1}, where it has been shown that it is not possible to obtain a realistic Higgs spectrum without including soft $A_4$-breaking terms. Indeed if we introduce a soft breaking part, eq. (\ref{softpot}), to the potential with adimensional parameters $m$, $n$, $k$ we can find five Higgses, all of which have masses in the LHC sensitive range between 100 GeV and 1 TeV. It is interesting to underline that such large Higgs masses have been recovered by using soft terms at most of order of $5\%$ of the EW vev. This underlines a non-linear dependence, as can be seen in fig. \ref{fig:soft}.

\begin{figure}[th!]
\begin{center}
\includegraphics[width=11cm]{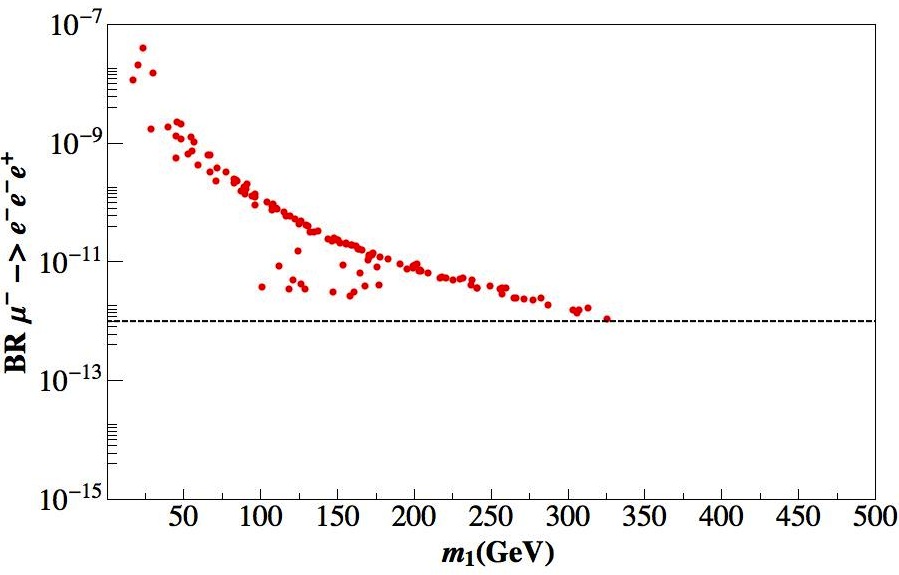}\\
\includegraphics[width=11cm]{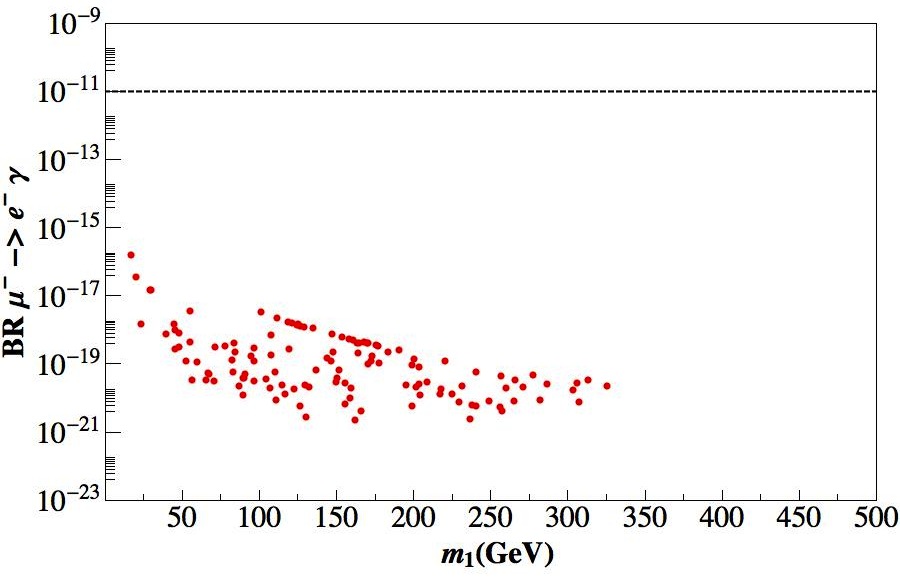}
\caption{\label{fig7}\it On the upper (lower) side, the branching ratio of the decay of a $\mu^-\rightarrow e^-e^-e^+$ ($\mu^- \rightarrow e^- \gamma$) versus the lightest Higgs mass. The horizontal band is the experimental limit \cite{PDG2010}.}
\end{center}
\end{figure}

In contrast with the Model 1, $A_4$ is completely broken by the vev of the Higgs triplet. As there is no residual symmetry, there are no special selection rules that forbid flavour changing interactions. In particular the processes $\mu^- \rightarrow e^- e^- e^+$ and $\mu^- \rightarrow e^- \gamma$ are allowed. The first process, fig. \ref{fig7} occurs at tree level and produces a strong bounds on the Higgs sector, where the lightest Higgs mass is expected to be above about 300 GeV.  On the other hand, the radiative muon decay to an electron, fig. \ref{fig7}, is loop suppressed and the new physics leads to a branching ratio below the observed experimental bound.

\subsection{Model 3}
\label{sec:Lav}

The Model 3 is built as an $A_4$ model for quarks \cite{LK:A4EWscale}, where both up- and down-type quarks couple to the Higgs triplet. There are eight parameters in their model whose values are unpredicted by the model itself, but are instead determined in order to reproduce the masses
of quarks and their mixing angles. The Yukawa matrices for both up and down quarks has the same form as that of charged leptons of the Model 1, given in eq. (\ref{yukawa_ma}). They provide then six parameters out of eight. The remaining two come from the vev of the triplet in the form $(e^{i \omega},e^{-i \omega},r)v_w/\sqrt{2+r^2}$, where $r$ is an adimensional quantity. Apart from a permutation in the three entries, this is the same vacuum used in Model 2.

The Higgs spectrum can only be realistic in the situation where $A_4$ is (softly) broken. Although $\omega$ is not absolutely constrained, the need of reproducing the neutrino mixing pattern suggests that the phase is small. In Model 2 we commented on the dependence of the Higgs masses on the soft parameters and the same applies in this case: the dependence is not linear and for even small soft parameters we get large Higgs masses. The plot in fig. \ref{fig:soft} is representative also of this model.

\begin{figure}[th!]
\begin{center}
\includegraphics[width=11cm]{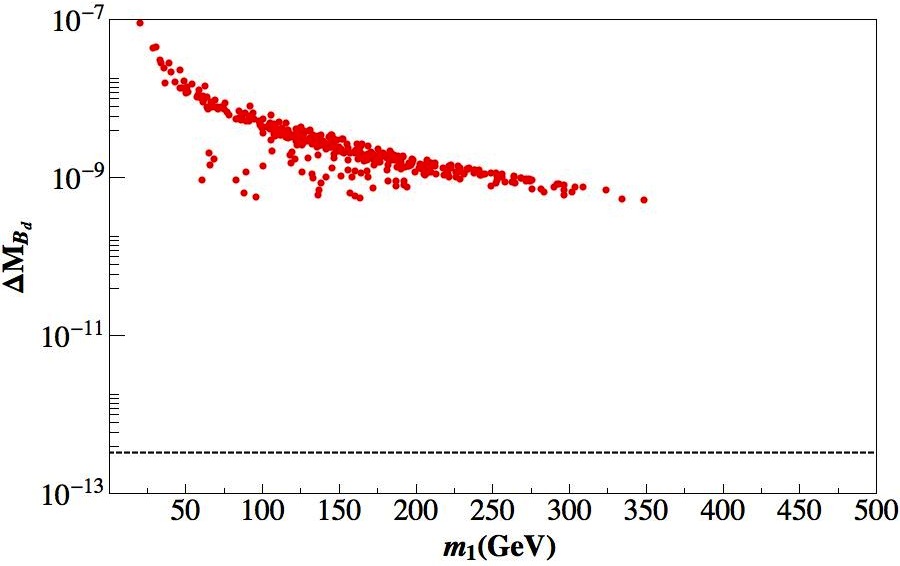}
\includegraphics[width=11cm]{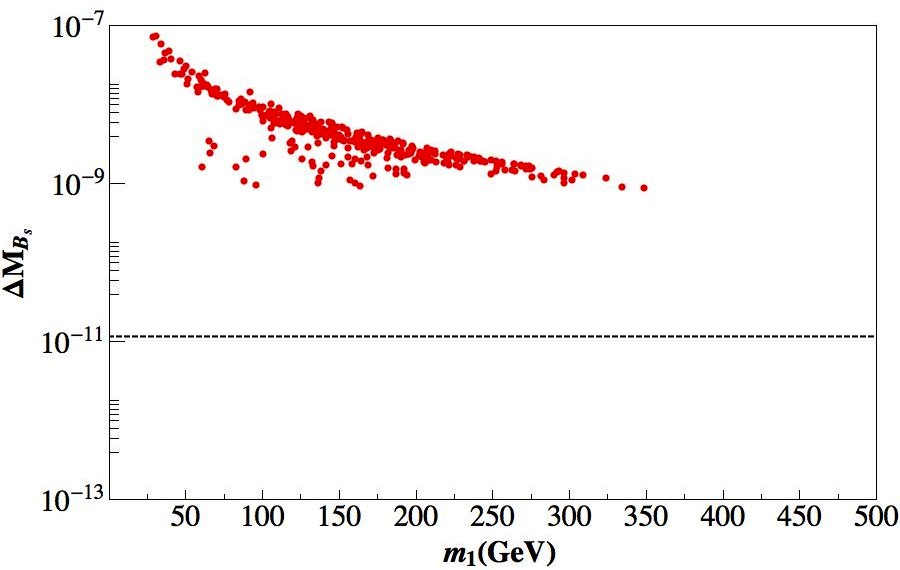}
\includegraphics[width=11cm]{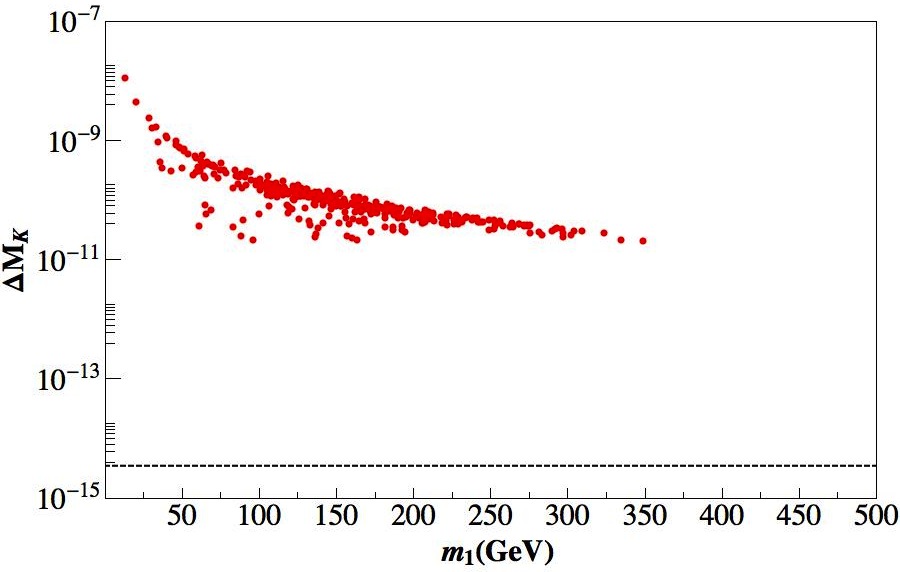}
\caption{\label{fig8}\it $\Delta M_F$ for $B_d$, $B_s$ and $K$ mass splittings versus the lightest Higgs mass in the Model 3. The horizontal lines correspond to the experimental values as reported in \cite{BCGI:HiggsFCNC}.}
\end{center}
\end{figure}

Experimentally, in the quark sector two features have been explored: flavour changing interactions and CP violation.
Remarkably, the CKM matrix obtained in the model under inspection is completely real. We will consequently focus only on flavour changing processes. As discussed in section \ref{sec:decay} meson oscillations are in these models mediated by tree level diagrams instead of box diagrams. We therefore expect strong bounds from the mass splittings in the neutral B-meson and Kaon systems. In fig. \ref{fig8}, we plot $\Delta M_F$ versus the lightest Higgs mass for these systems. Indeed $\Delta M_F$ is large, up to several orders of magnitude above the experimental value for the $B_d$ meson and the Kaon.

%
%

\section{Conclusions}

Flavour discrete symmetries are a popular tool to reproduce mass and mixing patterns of leptons and quarks. In a specific class of models,
the flavour scale is set to coincide with the EW symmetry by assigning the Higgs field to a non-trivial representation of the
underlying symmetry. In a previous paper \cite{ABMP:Constraining1} we analyzed the issues related to the construction and the viability of the potential in the scalar
sector, focusing on the choice of the non-Abelian discrete group $A_4$. We were able to discuss the minima of the potential and to set general constraints
to the framework without specifying the fermion content.
In this work we developed a very general formalism to describe the interaction of charged and neutral Higgs and of fermions. In the mass basis
of both Higgs bosons and fermions the interaction depends on the Yukawa matrices that appear in the Lagrangian and the unitary matrices that rotate
the flavour basis into the mass basis.

We applied the formalism to three specific models that implement the symmetry $A_4$. These models differ in the representations to which the fermions are
assigned and in the choice of the vacuum expectation values of the scalar fields.
The Model 1\cite{MR:A4EWscale} of lepton mixing has a vev in the direction $(v,v,v)$. In this setup, some transitions are
forbidden by the symmetry and the decay $\tau^- \rightarrow \mu^- \mu^- e^+$ becomes then the most relevant process. We studied its dependance
on the mass of the lightest Higgs and recognized that for the largest part of the assumed values the branching ratio is below the current experimental
limit.

Apart from a permutation of the components, both the Model 2 \cite{MP:A4EWscale} and Model 3 \cite{LK:A4EWscale} select the complex vev $(e^{i \omega},e^{-i \omega},r)v_w/\sqrt{2+r^2}$. The purpose of the two approaches is to reproduce lepton and quark masses and mixing, respectively.
The benchmark process in the lepton sector is the decay $\mu^- \rightarrow e^- e^- e^+$. Given the experimental bound, our analysis
showed that the Model 2 is disfavoured for values of the Higgs mass below 300 GeV. In the quark sector, $B$ and $K$ mesons oscillations mediated by Higgs exchange were considered and their predictions are largerly above the current experimental limit and strongly disproves the setup of Model 3.

In conclusion, we showed that a deep and careful analysis of the phenomenology of flavour models is fundamental to test their validity beyond the
prediction of the mixing patterns and is a powerful tool to discriminate among them.

%
%
\section*{Acknowledgments}

We thank Ferruccio Feruglio for interesting comments and discussions. The work of RdAT and FB is part of the research program of the Dutch Foundation for Fundamental Research of Matter (FOM). The work of FB has also been partially supported by the Dutch National Organization for Scientific Research (NWO). RdAT acknowledges the hospitality of the University of Padova, where part of this research was completed. AP recognizes that this work has been partly supported by the European Commission under contract MRTN-CT- 2006-035505 and by the European Programme �Unification in the LHC Era�, contract
PITN-GA-2009-237920 (UNILHC).

%
%

\newpage
\bibliography{ConstrainingHiggs2v5.bbl}

\begin{thebibliography}{10}

\bibitem{ABMP:Constraining1}
R.~de~Adelhart~Toorop, F.~Bazzocchi, L.~Merlo, and A.~Paris, arxiv: 1012.1791.

\bibitem{AF:ReviewDiscreteSyms}
G.~Altarelli and F.~Feruglio, Rev. Mod. Phys. {\bf 82} {(2010)} 2701--2729
  {[arXiv: 1002.0211]}.

\bibitem{MR:A4EWscale}
E.~Ma and G.~Rajasekaran, Phys. Rev. {\bf D64} {(2001)} 113012 {[arXiv:
  hep-ph/0106291]}.

\bibitem{MP:A4EWscale}
S.~Morisi and E.~Peinado, Phys. Rev. {\bf D80} {(2009)} 113011 {[arXiv:
  0910.4389]}.

\bibitem{LK:A4EWscale}
L.~Lavoura and H.~Kuhbock, Eur. Phys. J. {\bf C55} {(2008)} 303--308 {[arXiv:
  0711.0670]}.

\bibitem{HMPV:DiscreteDM}
M.~Hirsch, S.~Morisi, E.~Peinado, and J.~W.~F. Valle, {(2010)} {[arXiv:
  1007.0871]}.

\bibitem{MMP:DarkMatterA4}
D.~Meloni, S.~Morisi, and E.~Peinado, {(2010)} {[arXiv: 1011.1371]}.

\bibitem{PDG2010}
K.~Nakamura {\em et~al.}, J. Phys. {\bf G 37} {(2010)} 075021.

\bibitem{Lavoura:GenFormulae}
L.~Lavoura, Eur. Phys. J. {\bf C29} {(2003)} 191--195 {[arXiv:
  hep-ph/0302221]}.

\bibitem{MEGA}
M.~L. Brooks {\em et~al.}, Phys. Rev. Lett. {\bf 83} {(1999)} 1521--1524
  {[arXiv: hep-ex/9905013]}.

\bibitem{MEG}
A.~Maki, AIP Conf. Proc. {\bf 981} {(2008)} 363--365.

\bibitem{ARS:Phem2HDM}
D.~Atwood, L.~Reina, and A.~Soni, Phys. Rev. {\bf D55} {(1997)} 3156--3176
  {[arXiv: hep-ph/9609279]}.

\bibitem{Wells:HiggsLectures}
J.~D. Wells, arxiv: 0909.4541.

\bibitem{BCGI:HiggsFCNC}
A.~J. Buras, M.~V. Carlucci, S.~Gori, and G.~Isidori, JHEP {\bf 10} {(2010)}
  009 {[arXiv: 1005.5310]}.

\bibitem{Ma:SeeSawNONP}
E.~Ma, Phys. Rev. Lett. {\bf 86} {(2001)} 2502--2504 {[arXiv: hep-ph/0011121]}.

\bibitem{Ma:A4EWscale}
E.~Ma, Phys. Rev. {\bf D82} {(2010)} 037301 {[arXiv: 1006.3524]}.

\bibitem{FP:RareDecaysA4}
F.~Feruglio and A.~Paris, Nucl. Phys. {\bf B840} {(2010)} 405--423 {[arXiv:
  1005.5526]}.

\bibitem{Ma:DarkTB}
E.~Ma, Phys. Lett. {\bf B671} {(2009)} 366--368 {[arXiv: 0808.1729]}.

\bibitem{FHLM:LFVinA4}
F.~Feruglio, C.~Hagedorn, Y.~Lin, and L.~Merlo, Nucl. Phys. {\bf B809} {(2009)}
  218--243 {[arXiv: 0807.3160]}.

\bibitem{FHM:Vacuum}
F.~Feruglio, C.~Hagedorn, and L.~Merlo, JHEP {\bf 03} {(2010)} 084 {[arXiv:
  0910.4058]}.

\bibitem{FHLM:LFVinSUSYA4}
F.~Feruglio, C.~Hagedorn, Y.~Lin, and L.~Merlo, Nucl. Phys. {\bf B832} {(2009)}
  251--288 {[arXiv: 0911.3874]}.

\end{thebibliography}

\end{document}